\documentclass[useAMS,usenatbib]{mn2e}
\usepackage{graphicx}
\usepackage{latexsym}
\usepackage{xspace}

\title[Variable Accretion Rates and Fluffy First Stars]{Variable Accretion Rates and Fluffy First Stars}
\author[Smith, Hosokawa, Omukai, Glover \& Klessen]{Rowan J. Smith$^{1}$\thanks{Email: rowanjsmith.astro@googlemail.com}, Takashi Hosokawa$^{2}$, Kazuyuki Omukai$^3$, \newauthor Simon C.O. Glover$^1$, and Ralf S. Klessen$^1$\\
$^1$ Zentrum f\"ur Astronomie der Universit\"at Heidelberg, ITA, Albert-Ueberle-Str. 2, 69120 Heidelberg, Germany \\
$^2$ Jet Propulsion Laboratory, California Institute of Technology,
Pasadena CA 91109, USA \\
$^3$ Department of Physics, Kyoto University, Kyoto 606-8502, Japan
 }

\begin{document}

\pagerange{\pageref{firstpage}--\pageref{lastpage}} \pubyear{2010}

\maketitle

\label{firstpage}

\def\mnras{MNRAS}
\def\apj{ApJ}
\def\aap{A\&A}
\def\apjl{ApJL}
\def\apjs{ApJS}
\def\bain{BAIN}
\def\pasp{PASP}
\def\araa{ARA\&A}
\def\ga{\sim}
\def\nat{Nature}
\def\aj{AJ}
\def\pasj{PASJ}


\newcommand{\eq}{Equation }
\newcommand{\fig}{Figure }
\newcommand{\msun}{\,M$_{\odot}$}
\newcommand{\rsun}{\,R$_{\odot}$}
\newcommand{\gcmc}{\,g\,cm$^{-3}$}
\newcommand{\kms}{\,kms$^{-1}$}
\newcommand{\tab}{Table }
\newcommand{\gcms}{\,g\,cm$^{-2}$\xspace}
\newcommand{\E}{\times 10}
\newcommand{\msunyr}{\rm{M}_\odot~{\rm yr}^{-1}}
\def\spose#1{\hbox to 0pt{#1\hss}}
\newcommand\lsim{\mathrel{\spose{\lower 3.0pt\hbox{$\mathchar"218$}} \raise 2.0pt\hbox{$\mathchar"13C$}}}

\begin{abstract}
We combine the output of hydrodynamical simulations of Population III star cluster formation with stellar evolution models, and calculate the evolution of protostars experiencing variable mass accretion rates due to interactions within a massive disk. We find that the primordial protostars are extended `fluffy' objects for the bulk of their pre-main-sequence lifetimes. Accretion luminosity feedback from such objects is high, but as shown in previous work, has a minimal effect on the star cluster. The extended radii of the protostars, combined with the observation of close encounters in the simulations, suggests that mergers will occur in such systems. Furthermore, mass transfer between close protostellar binaries with extended radii could lead to massive tight binaries, which are a possible progenitor of gamma ray bursts.
\end{abstract}

\begin{keywords}
cosmology: early Universe -- stars: formation -- stars: pre-main-sequence
\end{keywords}

\section{Introduction}
It is becoming increasingly clear that Population III star formation is not as different from that in the present day as was once thought. Several authors have now shown that instead of a single massive star being formed at the centre of a primordial halo, a small cluster is formed (e.g.\ \citealt{Machida08}; \citeauthor{Clark11a}~2011a,b; \citealt{Turk09,Stacy10,Prieto11}).  There are frequent dynamical interactions within these clusters that can strongly alter the accretion rates of the protostars \citep{Greif11}. As such, Population III star formation begins to resemble simulations of local clustered star formation \citep[e.g][]{Bonnell11,Offner10,Girichidis11}. For a review of present day clustered star formation see \citet{MacLow04}.

A parallel strand in the study of Population III star formation has been the investigation of the stellar evolution of the protostar as it grows in mass.  As the protostar grows, the consequences of its protostellar evolution become increasingly significant. For example, during the protostellar stage of the evolution, radiation from the protostar alters the thermal and chemical structure of the circumstellar disk. As a result, the nature of disk fragmentation via the gravitational instability is affected by the protostellar radiation (\citeauthor{Clark11b}~2011b; \citealt{Smith11b}). As the star grows in mass and approaches the main sequence, it also starts to emit UV radiation. This creates an HII region which can prevent further mass accretion onto the star and limit the final stellar mass (\citealt{McKee08}; \citeauthor{Hosokawa11b}~2011b; \citealt{Stacy11c}). The stellar evolution also determines the  protostellar radius, which is an important quantity for understanding the dynamic evolution of a young cluster, as this determines the likelihood of protostellar collisions and mergers (for an analysis of present day protostellar clusters see \citealt{Baumgardt11}). 

Much of the pioneering work on the stellar evolution of Population III protostars was done by \citet{Stahler86a} and \citet{Omukai01,Omukai03} who numerically solved the 
interior structure equations assuming various constant accretion rates and showed that the evolution quantitatively changes with different rates.
However, as was pointed out in \citet[][hereafter S11]{Smith11b}, the protostars in a Population III minihalo do not accrete at a constant rate, but at a highly variable one, as the environment from which they accrete is not uniform. Time-variable mass accretion is also ubiquitous in present-day star formation \citep[e.g.][]{Schmeja04,Vorobyov06,Machida11} and its effects on protostellar evolution have begun to be explored (\citealt{Wuchterl01,Baraffe09,Baraffe10}; \citeauthor{Hosokawa11}~2011a; \citealt{Hartmann11}). In this paper, we seek to combine the strands of hydrodynamic simulations of fragmentation with stellar modelling and study how time-variable mass accretion affects the evolution of primordial protostars.

\section{Protostellar Accretion Rates}

We use protostellar accretion data taken from the simulations of S11. These are smoothed particle hydrodynamics (SPH) re-simulations of five cosmological minihalos presented in \citet{Greif11} in which the heating of the gas by the accretion luminosity produced by accreting `sink particles' \citep{Bate95,Jappsen05,Federrath10a} has been taken into account. Sink particles represent regions where protostars will form. They are non-gaseous particles that interact with their environment only through gravitational forces and are formed from bound and unambiguously collapsing fragments which are above a specified critical density, often chosen to be the density at which the Jeans length of the gas is only
just resolved. Through the representation of protostars as sink particles, the mass going into star formation, and the location of protostars can be tracked over the lifetime of the simulation. Relatively large sink radii of 20 AU were used in S11, which means that the inner disks surrounding our protostars are not resolved. However, by sacrificing the ability to follow these regions the computational time for the calculation is reduced sufficiently to allow the evolution of the protostars to be followed up to the point where they start contracting to the main sequence and produce substantial ionising radiation \citep{Tan08}. Our large sink particles also slightly suppress the total number of stars formed in our simulation, as we can no longer form fragments in the inner disk. These simulations, by having large sinks and by including accretion luminosity feedback, therefore represent a conservative estimate of the total amount of fragmentation that will occur within primordial halos.

\begin{figure}
\begin{center}
\includegraphics[width=3.3in]{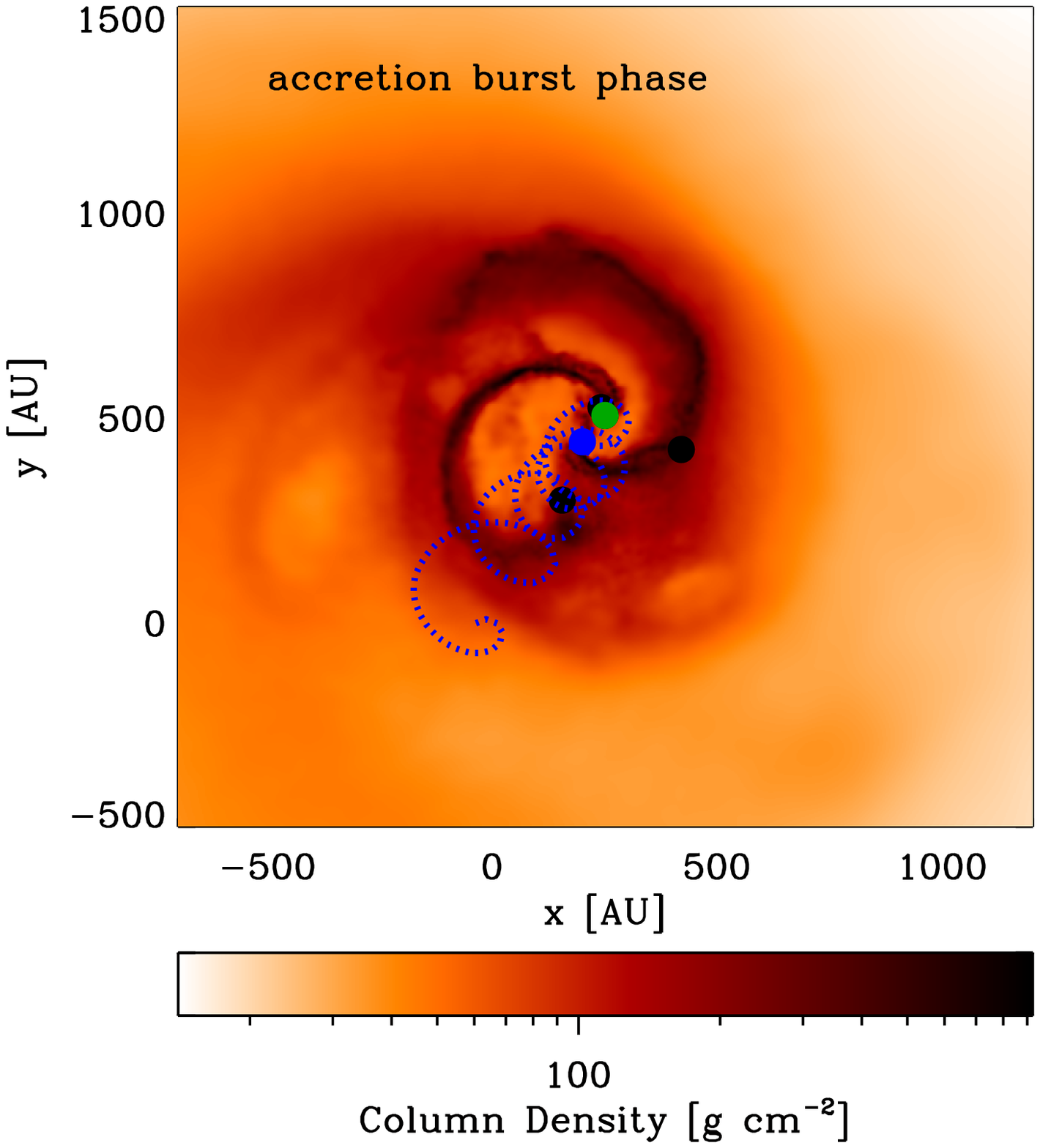}\\
\includegraphics[width=3.3in]{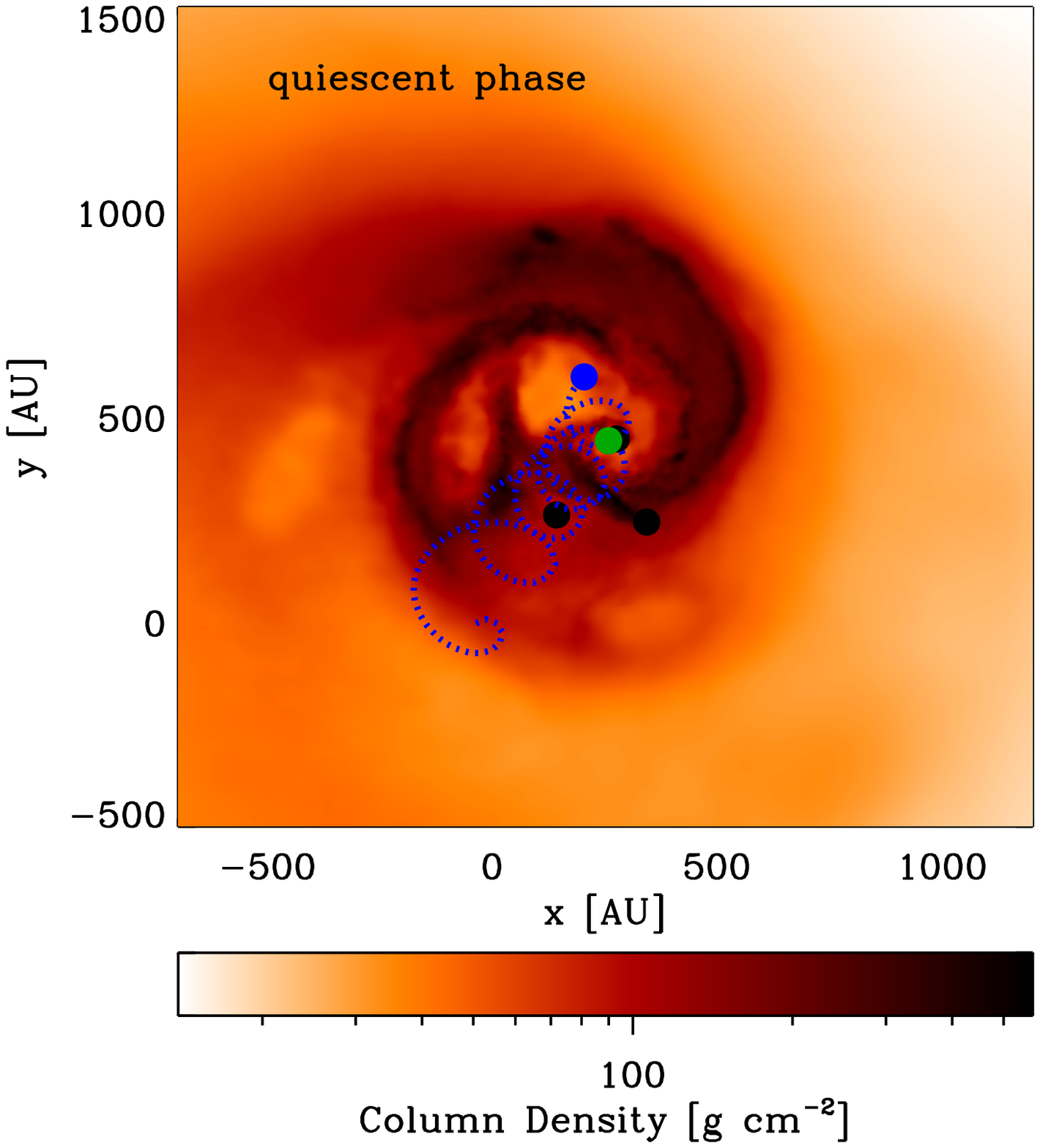}
\caption{A column density projection of the central regions of halo 4 from S11 where there is a gravitationally unstable disk from which multiple protostars are formed. Black circles show the location of sink particles representing protostars. Over-plotted on the figure is the path of one of the sink particles (shown in blue) which is orbiting within the disk. The central sink is shown in green. The top panel shows when the orbiting sink has a high accretion rate ($\dot{M} = 0.014$ M$_\odot$yr$^{-1}$, M$_*=10.35$ \msun) as it passes through the spiral arm close to the central sink and the bottom panel shows where it has a low accretion rate ($\dot{M} = 0.002$ M$_\odot$yr$^{-1}$, M$_*=10.45$ \msun) as it passes through a less dense region.}
\label{orbit}
\end{center}
\end{figure}

In \fig \ref{orbit} we show two snapshots of the central region of Halo 4 from S11. An extended thick disk is formed at the centre of the halo which has strong spiral arm features due to gravitational instability. \citeauthor{Clark11b}~(2011b) showed that primordial accretion disks become unstable due to the rate at which material can be transported inwards through the disk being lower than the rate at which new material falls on to the disk. This effect causes the disk in \fig \ref{orbit} to fragment into multiple additional sinks which orbit around the centre of mass of the system within the irregular accretion disk. The calculations in S11 had to be terminated when protostars had masses in excess of 20 \msun\ as they did not include the effects of ionisation feedback. However in reality the protostars will continue to accrete beyond this point, until the ionising radiation finally quenches the accretion 
(\citeauthor{Hosokawa11b}~2011b).

\begin{figure}
\begin{center}
\includegraphics[width=3.5in]{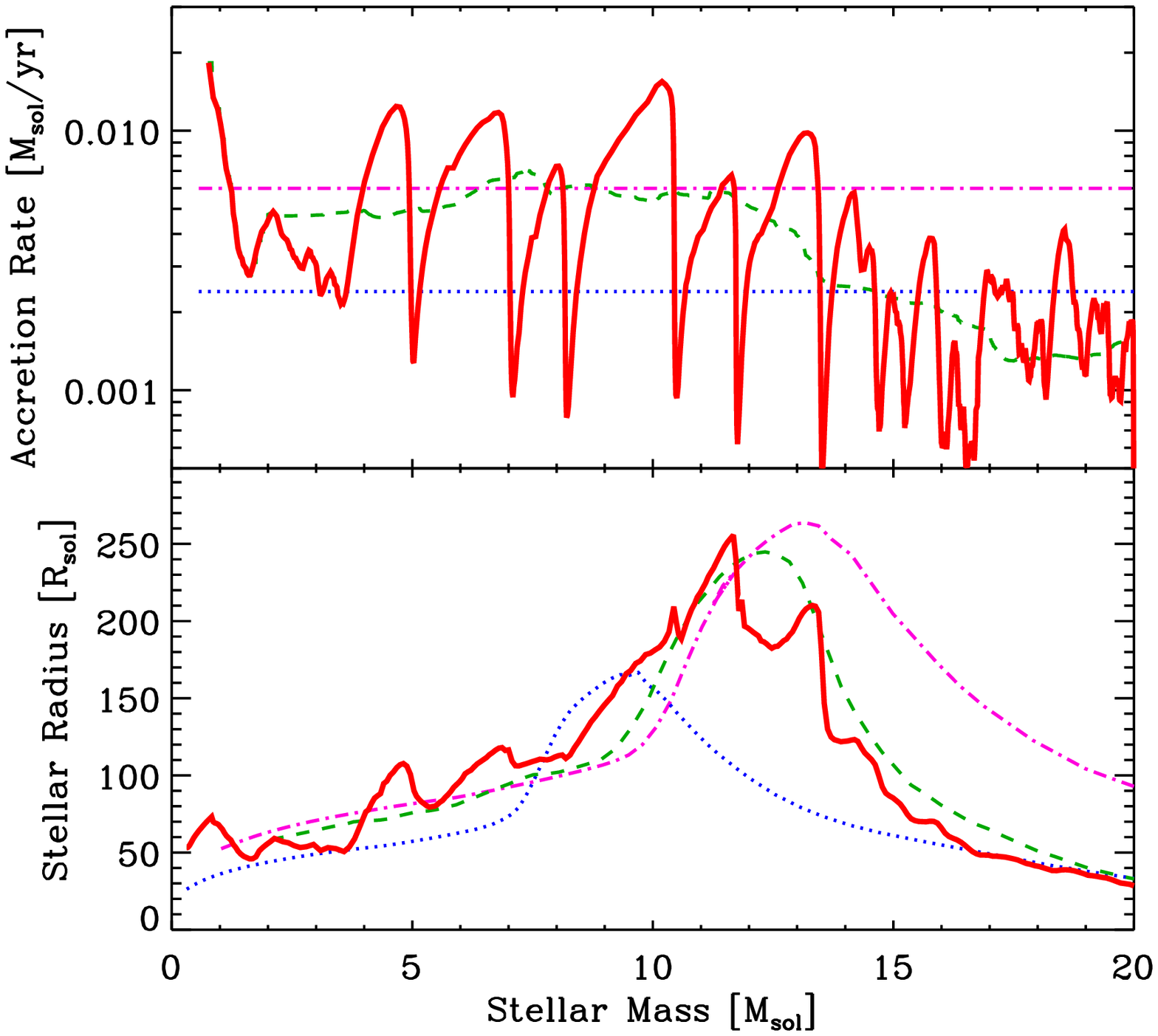}
\caption{\textit{Top}: The accretion rate of the orbiting sink in \fig \ref{orbit}. The solid red line shows the accretion rate measured directly from the simulation. The mean accretion rate at which 20\msun was accreted ($2.4\E^{-3}\, \msunyr$) is shown by the dotted blue line and a running average of the variable rate is shown by the dashed green line. An additional higher mean accretion rate of $6\E^{-3} \,\msunyr$ is plotted for comparison using a pink dot-dashed line. \textit{Bottom}: The evolution of the protostellar radius given the above accretion histories. The variable rate measured from the simulation results in the protostar having an extended stellar radius which is sustained over a large mass range. }
\label{radius4-3}
\end{center}
\end{figure}

The accretion rate of the orbiting sink in \fig \ref{orbit} is shown in the upper panel of \fig \ref{radius4-3}. The accretion rate is highly variable with sharp peaks and troughs. \fig \ref{orbit} shows that the sink orbits around the protostar at the centre of the disk. When the protostar is at the periastron of its orbit (as shown in the top panel of \fig \ref{orbit}) it passes though a dense one-armed spiral that causes the protostellar disk that surrounds the sink to grow in size, leading to a high accretion rate. On the other hand, when the protostar moves through the low density material outside the spiral arms (as shown in the bottom panel of \fig \ref{orbit}), its protostellar disk becomes depleted of gas, and the sink has a lower accretion rate. This variation leads to fluctuations of up to an order of magnitude in the accretion rate.

In our simulations we are missing the inner 20AU of the accretion disk as it is inside the sink particle radius. This raises the possibility that some of the variability in our measured accretion rates might have been smoothed by an inner accretion disk. However, as shown in \citeauthor{Clark11b}~(2011b), primordial accretion disks are massively self-gravitating and rapidly accreting even down to scales of 1.5 AU. Furthermore, the simulations in S11 did not use the instantaneous accretion rate, but instead used a smoothed average calculated over the last 100 years. This should to some extent mimic the effect of accreted material being buffered by the inner disk.

\section{Protostellar Evolution with Variable Accretion Rates} 
To calculate the stellar radii we use the measured variable accretion rates as input for a stellar evolution code, an approach first used in \citet{Wuchterl01}. We calculate the interior structure of the protostar by numerically solving the four stellar structure equations (i.e.\ the equations of continuity, hydrostatic equilibrium, energy conservation and energy transfer)  with the inclusion of mass accretion \citep{Omukai03, Hosokawa09, Hosokawa10}. 

The thermal efficiency of mass accretion, i.e., the specific entropy of the accreting material, is an unknown but important quantity which controls the effect of mass accretion on the structure of the protostar. Our numerical code handles the two opposite limits where the mass accretion is thermally efficient (``hot accretion'') and inefficient (``cold accretion'') by adopting  different outer boundary conditions (\citeauthor{Hosokawa11}~2011a). The exact value of the thermal efficiency of mass accretion should depend on the thermal structure of the accretion flow connecting the protostar and the disk. It is naively expected that the accretion flow becomes hotter for a higher accretion rate. Studies of FU Orionis outbursts in the Galaxy, for which mass  accretion at $\simeq 10^{-4}~\msunyr$ is expected, suggest that the gas temperature near the inner edge of the disk is $\simeq 10^4$~K \citep[e.g.,][]{Hartmann11}. The typical accretion rates in Population III star formation are $10^{-3}~\msunyr$ or more, and the disk temperature near the stellar surface should exceed $10^5$~K in such situations \citep{McKee08}. As such, hot accretion is the most realistic case for our model and we use this assumption in our following calculations.


\begin{figure}
\begin{center}
\includegraphics[width=3.5in]{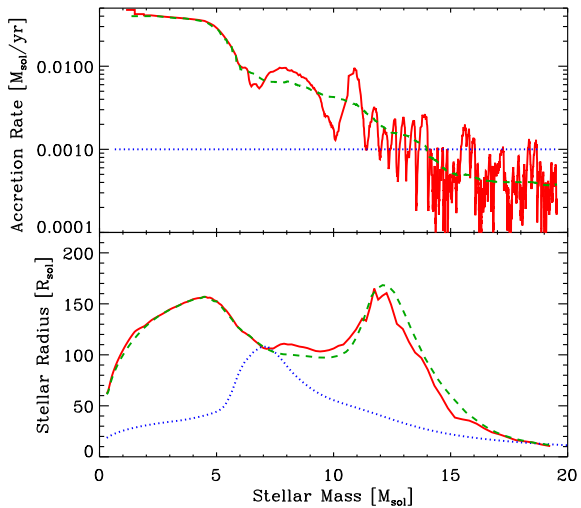}
\caption{As \fig \ref{radius4-3}, but for the central sink. In this case the mean accretion rate is $1\E^{-3} \msunyr$. The high initial accretion rate of the protostar, before fragmentation limits its accretion, causes the protostellar radius to be enhanced compared to the averaged rate.}
\label{radius4-1}
\end{center}
\end{figure}

Using our stellar evolution code we calculate the protostellar radius 
using four rates: the measured accretion rate from S11, a higher accretion rate, the mean 
accretion rate, and a running average of the measured rate. The mean 
accretion rate is found by dividing the final mass of the protostar in 
\fig \ref{radius4-3} by the protostar's age. The running average is 
calculated by averaging the measured accretion rate over a $\pm 3$
\msun\ mass range. 
This has the effect of smoothing out short period variations 
of $< 10^3$ years in the accretion rate.
Note that the simulations follow the evolution roughly over 
$10^4$ years.
The overall trend that the accretion rate decreases as the stellar mass increases is extracted in the running-average case.

The resulting stellar radii are shown in the bottom panel of \fig
\ref{radius4-3}. The variable accretion rate has increased the radius of
the protostar compared to the mean rate. The stellar radius now extends
to a maximum value of over 200\rsun, or roughly 1AU. The maximum value
of the protostellar radius before it reaches the main sequence is
similar to what we find when we use a higher constant accretion rate
of $6 \times 10^{-3}~\msunyr$. This is unsurprising, as we see that the
running average of the measured accretion rate is roughly equal to 
$6 \times 10^{-3}~\msunyr$ for protostellar masses $M_* < 12~M_\odot$.
For larger $M_{*}$, the running-average of the accretion rate decreases
significantly, eventually becoming comparable to the mean accretion rate.
Because of this decrease, the use of the higher constant accretion rate
does not yield an accurate prediction for the protostellar radius at late times.
After the protostar starts contracting to the main sequence, the stellar 
radii of the mean, running-average and full models converge. 
Consequently, the end state of the protostar is set by the mean accretion 
rate and is not affected by the earlier variability in its accretion history. 



As the variation shown for the protostar in \fig \ref{radius4-3} is an extreme case, we also repeat this calculation with the protostar at the centre of the disk in \fig \ref{orbit}. \fig \ref{radius4-1} shows that once again the protostellar radius is substantially enhanced relative to the mean rate. The accretion rate of the central protostar drops rapidly after it reaches a mass of around 6\msun. This corresponds to the point in our simulation where the disk around the star becomes unstable and the additional protostars in the halo are formed. The new protostars also accrete gas from the massive central disk and hence reduce the accretion rate of the central object in a process termed fragmentation-induced starvation \citep{Peters10c}. The evolution of this protostar's radius cannot be reproduced by any model with a constant accretion rate as it has two peaks in its evolution and a constant accretion rate can only produce one.

\begin{table}
	\centering
	\caption{The peak and mean radii for the measured variable accretion rates and the constant mean rate.}
		\begin{tabular}{l c c }
   	         \hline
	         \hline
	          & Variable rate & Mean rate \\
	         \hline
	         \textbf{Orbiting protostar} & &\\
	         Peak Radius  & 255.7 \rsun  & 166.7\rsun \\
	         Mean Radius  &  102.5\rsun & 63.4\rsun \\
		\hline	         
		\textbf{Central protostar} & &\\
	         Peak Radius  & 165.0\rsun & 108.8\rsun\\
	         Mean Radius  &  107.5\rsun & 35.9\rsun\\
   	         \hline
	         \hline	         
		\end{tabular}
	\label{peak}
\end{table}

\tab \ref{peak} shows the maximum and average value of the stellar radius of each sink for the variable accretion rate and its mean. In both cases the protostellar radius has a higher maximum and a higher mean value when the variable rate is used. In particular, for the central sink shown in \fig \ref{radius4-1} the mean protostellar radius is three times higher than that found using the mean accretion rate.


We find that the young protostars have large initial radii due to having an enhanced rate of accretion during the early phase of their evolution relative to that in later stages. In \fig \ref{radius4-3} the accretion rate of the sink is at an initially high level due to the sink accreting large amounts of material from the spiral arms within the dense disk. In \fig \ref{radius4-1} the rate was initially high but the formation of protostars in a disk around the central protostar caused the accretion rate in the later stages to fall due to fragmentation-induced starvation. Additional dynamical ejections can also remove sinks from the dense central regions of the halo and cause their accretion rates to fall, which once again means that the accretion occurs disproportionately in the early stages of the protostars' evolution.

The importance of the initial accretion rates can be seen by considering
the phases of stellar evolution. 
While the protostellar mass is less than around 10\msun, the
protostellar evolution is adiabatic, as the accretion timescale for 
the protostar (i.e., the timescale on which it increases its mass) 
is shorter than the Kelvin-Helmholtz (KH) timescale (i.e., the timescale on which
the star can redistribute entropy internally)
\begin{equation}
t_{\rm KH} = \frac{G M_*^2}{R_* L_*},
\end{equation}
where $M_*$ is the protostar's mass, $R_*$ the radius and $L_*$ the
luminosity.
Rapid accretion during this stage dumps large amounts of entropy onto 
the surface of the protostar, which cannot be quickly redistributed, 
causing the stellar surface to swell. 
The opacity in the stellar interior decreases as 
the stellar mass increases, which in turn causes the stellar luminosity to increase
\citep[e.g.,][]{Stahler86a}.
The KH timescale then becomes shorter and shorter, and finally falls below
the accretion timescale. 
After that point ($M_* \simeq 12$\msun\ in both the cases), 
the protostar begins to contract, losing its energy via radiation. 
The characteristic timescale of the evolution is the accretion timescale
in the early adiabatic accretion stage and the KH timescale in the
later contraction stage. Both timescales  are around $10^3$ years.

Using the running average of the accretion rate allows us to investigate
whether the short-term variability or the long-term trend in the accretion rate is
the greater factor in determining the stellar evolution. In Figures
\ref{radius4-3} and \ref{radius4-1}, the protostellar radius calculated
from the running average of the accretion rate provides an excellent
match to the true value. Consequently, the long-term trend of the 
accretion is more important than the short-term variability. 
It is reasonable that accretion variability shorter than the
characteristic timescale of $\sim 10^3$ years does not significantly influence the
overall evolution of the protostar.

Generally protostellar accretion rates are predicted to fall exponentially with time \citep[e.g.][]{Schmeja04}, but for the orbiting protostar the accretion rate is maintained at a high value throughout its early evolution which increases its protostellar radius. Conversely, the accretion rate experienced by the central protostar rapidly decreases after the disk fragments but during the adiabatic stage of its evolution the accretion rate is very high (over $0.01\, \msunyr$) and consequently it has a large initial stellar radius. The process of disk fragmentation therefore leads to a system in which the protostars have extended radii.

\citet{Hosokawa09} have shown that all massive protostars accreting at high rates are fluffy extended objects, but the Population III stars here are particularly striking examples. While the final main sequence radius of the stars considered here is small, the large radius during the protostellar stage has ramifications for the type of stellar cluster that is formed within the primordial minihalo. This is due to the accretion rate and stellar radius of the protostar determining both the strength of protostellar feedback in the young cluster, and the likelihood of stellar interactions and mergers, as we shall discuss in the following sections.

\section{Feedback}\label{feedback}

\begin{figure}
\begin{center}
\includegraphics[width=3.5in]{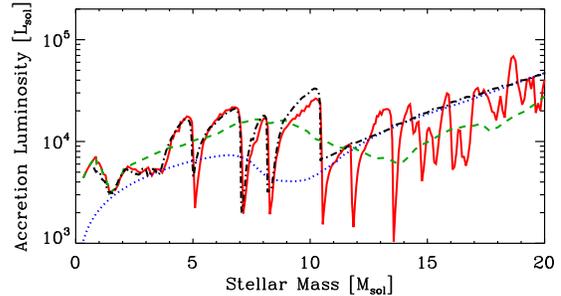}
\caption{A comparison of the accretion luminosity from the previously considered orbiting sink. The solid red curve uses the variable accretion rate, the dotted blue curve the constant rate and the dashed green curve is the running average rate. The black dashed-dotted curve shows the accretion luminosity calculated using the semi-analytic method used in S11.}
\label{lacc}
\end{center}
\end{figure}

The dominant form of stellar feedback during the first few thousand years of a protostar's life in a primordial halo is that from accretion luminosity. The luminosity is given by
\begin{equation}
L_{\rm acc}=\frac{GM_*\dot{M}}{R_*},
\end{equation} 
where $\dot{M}$ is the accretion rate.
The accretion rate contributes doubly to this relationship, as in addition to the direct dependence of $L_{\rm acc}$ on $\dot{M}$, the accretion rate also affects the radius of the star. \fig \ref{lacc} shows the accretion luminosity of the previously considered orbiting sink calculated with the measured accretion rate, the constant rate and the decreasing accretion rate. During the early evolution of the protostar the variable accretion rate gives accretion luminosities above $10^4$ L$_\odot$, which is substantially higher than the value we find for the constant accretion rate. S11 considered the effects of accretion luminosity in detail using a semi-analytic model based on the work of \citet{Omukai03} for radii and accretion luminosities. In this semi-analytic model the stellar radius evolves as
\begin{equation}\label{imfeq}
R_* \propto \left\{
    \begin{array}{ll}
         26 M_*^{0.27}(\dot{M}/10^{-3})^{0.41} &  M_*\leq p_1 \\
         A_1 M_*^{3} & p_1 \leq M_* <p_2 \\
         A _2 M_*^{-2} & p_2 \leq M_* \mbox{ \& } R< R_{ms} \\
    \end{array}
\right.
\end{equation}
where $R_*$ has units of R$_{\odot}$, $M_{*}$ has units of M$_{\odot}$, and the accretion rate $\dot{M}$ has
units of M$_{\odot} \: {\rm yr^{-1}}$. In this expression, the transition points between the evolutionary phases, 
$p_1$ and $p_2$, scale as
\begin{equation}
    \begin{array}{l}
	p_1=5 (\dot{M}/10^{-3})^{0.27}  \mbox{	\msun}, \\
	p_2=7(\dot{M}/10^{-3})^{0.27} \mbox{	\msun}, 
     \end{array}
\end{equation}
and the main sequence radius is given by
\begin{equation}
R_{\rm ms}=0.28 M_*^{0.61}  \mbox{	\rsun}.
\end{equation}
The accretion luminosity is then calculated from the measured accretion rate and the stellar radius. This model is plotted in \fig \ref{lacc} in black, and is a good match to the true accretion luminosity, particularly during the early stages of the protostar's evolution where the star evolves adiabatically. We therefore find that the conclusion of S11 that accretion luminosity does not prevent fragmentation of primordial gas is valid, despite the fact that S11 used a simplified semi-analytic approach rather than a full stellar evolution model.

As shown in \citeauthor{Hosokawa11b}~(2011b), ionising radiation plays a key role in determining the final mass of a Population III protostar. We find that the protostar reaches the main sequence at the same stage with either a constant or variable accretion rate, and therefore in both cases the protostar starts ionising its surroundings at the same mass.

\section{Close Encounters}
As the protostars are formed from disk fragmentation, the separation between sink particles is necessarily small. In fact, close encounters seem to be a feature of the small clusters formed in Population III halos. In the study of \citet{Greif11}, which considered fragmentation on scales down to 100 \rsun, many encounters were seen where two sink particles came within this radius. In the simulations considered here, we do not resolve fragmentation at separations of less than 20 AU and so do not follow the evolution of fragments formed in the inner parts of protostellar accretion disks. We therefore see fewer of these close encounters. Despite this, in all but one of the five halos studied in S11 there is at least one pair of sink particles which come within a distance of 1 AU of each other.

\begin{figure}
\begin{center}
\includegraphics[width=3.5in]{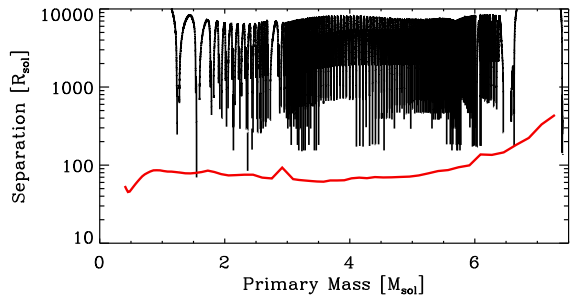}
\caption{The separation between the sink particles in the tightest binary seen in our hydrodynamic simulations (black). The red line shows the total stellar radius of the protostars during the evolution. As the radii of the protostars is frequently comparable to their separation these particular sinks are no longer well represented by point masses in this extreme case. }
\label{sep}
\end{center}
\end{figure}

As an illustrative example, in \fig \ref{sep} we plot the distance between two sinks that make up a tight binary in halo 3 of S11. This binary is the closest found in the simulations carried out by S11. On the same graph we also plot the total stellar radii of the two protostars, calculated from our stellar evolution code given the measured accretion rates of the sink particles. The binary orbit of the two sink particles is a very close one, and their separation is frequently comparable to their total radius. In our original simulations, the sinks were treated as point masses once formed. This is still a good approximation for the majority of the sinks in our simulations which do not have such close encounters. However, for extreme cases such as that shown in \fig \ref{sep}, this assumption clearly breaks down.

The tidal radius for a fluid satellite was calculated by Roche in 1847 to be
\begin{equation}
r_t\approx 2.44 R_p \left( \frac{\rho_p}{\rho_s}\right) ^{1/3},
\end{equation}
where $r_t$ is the tidal radius, $R_p$ the primary radius, $\rho_p$ the primary density and $\rho_s$ is the density of the secondary. The protostars considered here are well approximated as fluids as the outer extended layers will be easily separated from their parent protostar. Protostars which have encounters are usually formed close to each other and have similar masses, radii and accretion histories. Consequently their internal densities are also likely to be similar. Therefore we should expect tidal disruption of protostars which come within a distance of about 2.5 times the primary radius. To estimate the number of protostars in each halo likely to be affected by tidal effects we calculate the number of protostars which come within a distance of 3 primary radii of each other (we round up to the nearest integer to account for the effects of the secondary mass and oblateness). Since we do not have the full stellar evolution calculation for each star in our halo we use the radius from the semi-analytic model used in S11 which we introduced in Section \ref{feedback}.

\begin{table}
	\centering
	\caption{The number of sink pairs where the sinks pass within each other's tidal radius, compared to the total number of sink particles formed in each halo simulated in S11.}
		\begin{tabular}{l c c c c c }
   	         \hline
	         \hline
	           Halo &1 & 2 & 3 & 4 & 5 \\
	         \hline
	         Within Tidal Radius & 3/11 & 0/8 & 2/6 & 1/6 & 5/23 \\
	         Radii Touching  &  2/11 & 0/8 & 2/6 & 1/6 & 1/23  \\
   	         \hline
	         \hline	         
		\end{tabular}
	\label{encounters}
\end{table}
\tab \ref{encounters} shows the results. In all of the halos, bar one, there will be tidal disruption or mass transfer between a few protostars. A comparison to local star formation which might be of use to understand the physics of such an encounter is the recent work of \citet{Lajoie11} in which mass transfer is seen between an approximately equal mass binary on an elliptical orbit after the system has reached periastron. We also calculate in how many cases the stellar radii go on to touch (see \tab \ref{encounters}), and see that this is also common. We can therefore conclude that mergers will contribute to the growth of some Population III stars, although they will not necessarily be the dominant mode of protostellar growth.

Protostars that form a binary without merging may undergo mass transfer via Roche lobe overflow. This will influence their subsequent evolution and orbital separation. \citet{Krumholz07c} examine such an evolution in the context of present-day massive star formation, where the protostars have extended radii like those seen here due to high accretion rates. They show that in many cases the binary stellar mass ratio approaches unity and the orbital separation shrinks owing to the rapid mass exchange. The final outcome is a tight ($<$ 1AU) massive binary consisting of two almost identical stars. Such `massive twins' are a common occurrence in the local universe \citep[e.g.][]{Hilditch05}, and if this were also true in primordial star formation, they would be strong candidates for the progenitors of gamma-ray bursts (GRBs) in the early universe \citep[e.g.][]{Bromm06}. \citet{Fryer99} have shown that evolved binary systems consisting of compact objects, such as neutron stars or black holes, can lead to GRBs from accretion \citep{Woosley93}, or through mergers. Such systems could also form X-ray binaries during Roche lobe overflow from a post-main-sequence secondary on to a compact primary, and in the case of mergers may even be a source of gravitational waves \citep{Fryer99}.


While not all protostars will merge or undergo mass transfer, their extended fluffy structure introduces additional degrees of freedom to the stellar system during dynamical interactions. This means that energy can be dissipated during close encounters by tidal distortion of the surfaces of the protostars. As a result, the post-encounter velocities of protostars that undergo close encounters may be smaller than the values computed in current simulations in which the sink particle is approximated as a solid body. Future studies of fragmentation in primordial halos should attempt to include these effects to obtain the most accurate description of the growth of Population III protostars and their final masses. Recently \citet{Greif12} made the first step in this direction in a model of the first 10 yr of a protostellar disks evolution.

\section{Effects of Cold Mass Accretion}

\begin{figure}
\begin{center}
\includegraphics[width=3.5in]{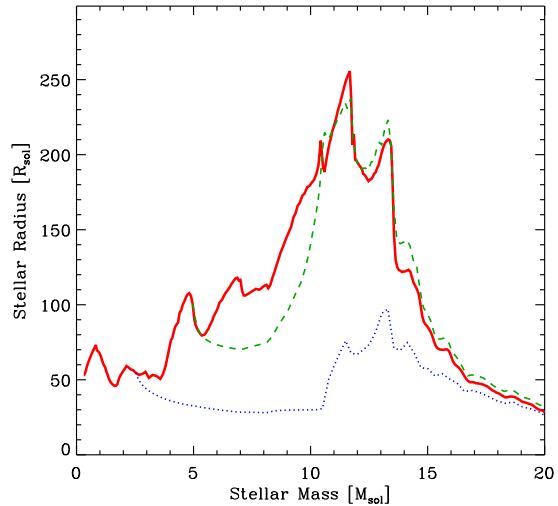}
\caption{
Effects of cold (low-entropy) mass accretion on 
the evolution of the stellar radius with the same variable 
accretion history as in Figure 2. 
The blue and green lines represent the cases where 
cold accretion is adopted after the stellar mass exceeds 
$5~M_\odot$ and $2.5~M_\odot$. 
The red line represents the fiducial case, where hot accretion is assumed throughout the evolution.
}
\label{coldacc}
\end{center}
\end{figure}

So far we have focused on the evolution of protostars undergoing thermally efficient mass accretion (``hot accretion''). As discussed in Section 3, hot accretion is the most appropriate model for the rapid mass accretion expected in Population III star formation. However, for a better understanding of the protostellar evolution, it is useful to examine the effect of reducing the thermal efficiency of the accretion. Experiments of this kind were presented in \citeauthor{Hosokawa11}~(2011a), where we examined the potential effects of different thermal efficiencies by switching the outer boundary conditions at some point during the evolution. Figure~\ref{coldacc} shows the evolution of the stellar radius in such cases, taking the same variable accretion history as in Figure 2.

The evolution depends critically on the protostellar mass, $M_*$, at the point at which the boundary conditions are switched. With the cold accretion for $M_* > 5~M_\odot$, the maximum stellar radius is almost the same as in the fiducial hot accretion case. When the boundary condition is switched at $M_* = 2.5~M_\odot$, the protostar initially contracts and then re-expands to a radius of $\simeq 100~R_\odot$ at a mass of around 10~M$_\odot$. This indicates that the cold mass accretion reduces the maximum stellar radius. The explanation for this phenomena comes from the stellar radius being at its maximum just prior to the KH contraction stage. In this transitional phase, heat accumulated in the stellar interior is transported outward and escapes from the star. A gas layer near the stellar surface will receive a portion of this transported entropy, which makes the star inflate \citep[e.g.,][]{Stahler86a}. This effect would be alleviated if the mass accretion was thermally inefficient and the average entropy in the stellar interior was reduced.

\section{Conclusions}
We have combined the accretion rates measured from Population III protostars in hydrodynamic simulations with a stellar evolution code to determine the structure of the protostars. The measured accretion rates had a large degree of variability and were particularly high during the adiabatic stage of the protostars' evolution. This increased the stellar radius of the protostars, causing them to become puffy extended objects. In the early stages of the protostar's life this intermittently increased the accretion rate of the protostar, but as shown in \citet{Smith11b} this has only a small effect on the dynamic evolution of the halo. In some cases the separation of the protostars during their dynamic evolution was sufficiently close that mass transfer and mergers may occur between protostars. Mass transfer between massive protostellar systems could lead to approximately equal-mass massive binaries that are a possible precursor of GRBs. The extended size and relatively large tidal radii of Population III protostars implies that interactions between them are likely to dissipate more energy than if they were pure point masses.

\section*{Acknowledgements}

We would like to thank Volker Bromm, Paul Clark and Naoki Yoshida for stimulating discussion, and thank the anonymous referee for useful feedback on the first version of this paper.  R.J.S and R.S.K.\ acknowledge support from the DFG via the SPP 1573 {\em Physics of the ISM} (grants SM321/1-1 \& KL 1358/14-1). R.J.S.\ also acknowledges the support of a Frontier grant of Heidelberg University sponsored by the German Excellence Initiative and the Landesstiftung Baden-W\"urttemberg under research contract P-LS-SPll/18 ({em Internationale Spitzenforschung ll}). T.H.\ appreciates the support by Fellowship of the Japan Society for the Promotion of Science for Research Abroad. K.O.\ acknowledges the support by the Grants-in-Aid from the Ministry of Education, Culture, and Science of Japan (2168407 and 21244021).

\bibliography{Bibliography}
\label{lastpage}

\end{document}